\documentclass[amsmath,amssymb,twocolumn,nofootinbib]{revtex4}
\usepackage{latexsym,comment,verbatim}
\usepackage{amssymb,multirow}
\usepackage{amsfonts}
\usepackage{amsmath,color}
\usepackage{graphicx}
\usepackage{bm}

\def\ber{\begin{eqnarray}}
\def\eer{\end{eqnarray}}
\def\beq{\begin{equation}}
\def\eeq{\end{equation}}

\def\ed{\end{document}}

\begin{document}

\title{Online Tutoring in Introductory Physics Courses: a Lockdown Experience}

\author{Matteo Luca Ruggiero}
\email{matteo.ruggiero@polito.it}
\affiliation{Politecnico di Torino, Corso Duca degli Abruzzi 24, Torino, Italy}

\author{Lorenzo Galante}
\email{lorenzo.galante@polito.it} % optional
\affiliation{DISAT \& TLLAB, Politecnico di Torino, Corso Duca degli Abruzzi 24, Torino, IT}
\affiliation{INFN Torino, Via Pietro Giuria 1, Torino, IT}
\affiliation{Enrico Fermi Historical Museum of Physics and Study and Research Centre, Piazza del Viminale, 1, Roma, IT}

\date{\today}

\begin{abstract}

Social distancing due to the Covid-19 pandemic deeply impacted on education worldwide, since schools and universities had to rapidly organise lessons and courses on line. Traditional interactions between teachers and students and, also among students, had to change and was substituted by on line connections. In this context, laboratory work and tutoring, which have an important role in the peer instruction model, needed to be redesigned. Here, we discuss an on line tutoring model adopted for the introductory physics courses at Turin Polytechinc University and evaluate its effectiveness by analysing the students performance both during the semester and the summer and autumn exam sessions.

\end{abstract}

\maketitle

%------------------------Section-------------------------
\section{Introduction}\label{sec:intro}
%------------------------Section-------------------------

The impact of the ongoing Covid-19 pandemic on education is without precedents since, overnight,  it deeply changed  teaching methods and organisations for billions of students and teachers  around the world. Schools and universities had (and still have)  gigantic challenges  to adapt their classes to distance learning as quickly as possible, in order to keep educational continuity.

Besides the obvious difficulties - both for teachers and students - due to the switch in a very short time to  online courses, there was an additional complication for the learning process, provoked by the atypical situation \cite{o2020feynman,tan2020bringing,dew2020student,klein2020studying}. In fact, because of  social distancing, interactions with teachers and peers considerably changed,  and, \textit{de facto}, students were forced to experience independent study much more than in the past \cite{moore1973toward}. Accordingly, the role of teachers and students became different from those  in a traditional learning process:  in this case teachers cannot directly interact with students and just act  as facilitators to support them, while students have to independently develop their collaborative efforts \cite{dietrich2020attempts}. Fortunately, due to the development of the Internet and related technologies, today's students have  devices such as smartphones, tablets and personal computers available. These technolgoies make it possible and easy to access large volumes of information and to maintain contact with classmates and educators. Undoubtedly, if this pandemic  had spread only 30 years ago, its effect on education systems worldwide would have been devastating.  

In addition, activities such as laboratory work and tutoring were greatly limited during the pandemic and they had to be  re-designed in order  to be effective in the learning process. These activities have an important role in the peer instruction model, whose effectiveness is well known, from  early childhood education \cite{damon1984peer}  up to college \cite{lasry2008peer} and university level \cite{mazur2013peer,crouch2001peer}.  In the literature \textit{peer collaboration} refers to laboratory work and  \textit{peer tutoring} to a tutoring model in which advanced students (``student-tutors'') or those in later years take on an instructional role (see e.g. \citet{boud2014peer,collings2014impact}), mainly aimed at the revision of the taught content. Peer  tutoring has several advantages, as reported by \citet{sneddon1759attitudes}: for instance, students are freer to communicate with each other in the absence of teachers; student-tutors have an approach to the course materials and resources which is more similar to the student one; the interaction between students and student-tutors is more direct and personal which can lead to a more lively and open learning environment. Usually, in this model there are meetings two/three times per week, in which one or more student-tutors works with a small number (ranging from 50-100) of students.  There is evidence that in this model both the tutees and the tutors get benefits \cite{sneddon1759attitudes}.

Social distancing provoked by the pandemic acts as a serious deterrent for carrying out peer tutoring which is typically intended for weak and less experienced students. For these reasons it is necessary to completely redesign this model of tutoring, and the same, of course, holds true for laboratory work \cite{pols2020physics,bradbury2020pandemic}. 

In this paper we report the results of an on line tutoring model developed for the introductory physics courses at Turin Polytechinc University during the lockdown due to the Covid-19 pandemic. The tutoring was  based on online activities, both synchronous, such as video  and chat sessions, and asynchronous, such as questionnaires and exercises. The former were conducted by student-tutors and organised by staff members. These activities  aimed to facilitate the  students' learning process and, also,  to mimic the classroom interactions that were not allowed.  In this work we aim to evaluate the effectiveness of the tutoring model by analysing the performances of the students involved, both during its progress and during their exams.

The paper is organised as follows: in Section \ref{sec:model} we describe the tutoring model and the context, while in Section \ref{sec:res} we examine the methodologies and give the results, which are then discussed in Section \ref{sec:disc}. Conclusions are eventually drawn in Section \ref{sec:conc}.

%------------------------Section-------------------------
\section{The Online Tutoring Model}\label{sec:model}
%------------------------Section-------------------------

Turin Polytechinc University (TPU) is one of the most important Italian Technical Schools for Engineers and it attracts students not only from Italy, but also from more than 100 countries  around the world;  about 45\% of students come from outside  Piemonte, the region where Turin is located, and 15\% come from abroad. About 5000 students  start studying at  TPU every year, and during the first year they are divided into 20 courses: 18 are taught in Italian and 2 in English. The students in the Italian courses are divided by alphabetical order rather than  ability.  During the first year, students attend the "Physics I" course, which is an introductory calculus-based course including classical mechanics and thermodynamics, Newtonian gravitation and  basic electrostatics in which Coulomb's  and Newton's law are compared.  In each  course,   there are between 300 and 500 students: besides first year students, there are also students who did not succeed in passing the exam from previous years. As a consequence, each year there are approximately 10000 students  in the Physics I courses.

The Physics I course is generally considered as difficult  and among those which most frequently contributes  to  dropping out of TPU. Consequently,    TPU organises tutoring activities to help students, in the spirit of the Learning Assistants scheme, developed by the University of Colorado \cite{otero2010physics}, where experienced students (usually 2 years older) are employed as tutors and, together with course lecturers, they revise the key topics. Usually, student-tutors work in pairs, and meet  small groups of students once or twice per week. The interaction is direct and immediate and mainly involves solving exercises.

The Physics I course is taught during the second semester of the academic year, which usually starts between the end of February and the beginning of March, so  in 2020 overlapped with the outbreak of the Covid-19 pandemic in Italy. Due to the emergency situation, all courses were switched to online delivery at TPU and the usual tutoring in conditions of social distancing could no longer take place; consequently, it was necessary to  design a new online tutoring model.

The model we developed is based on  a Moodle platform, integrated in the Internet TPU portal, which is accessible to all Physics I students; 20 student-tutors (here and henceforth \textit{junior tutors}) were recruited and coordinated by 4 TPU staff members (here and henceforth \textit{senior tutors}), who were lecturers in the Physics I courses.  The junior tutors were divided into 10 pairs and each pair was assigned to two courses. The activities described here refer to the  courses taught in Italian only because, due to the very short time available to organise the tutoring, it was not possible to involve  the 2 courses taught in English.

\begin{widetext}

\begin{figure}[ht]
\begin{center}
\includegraphics[scale=.35]{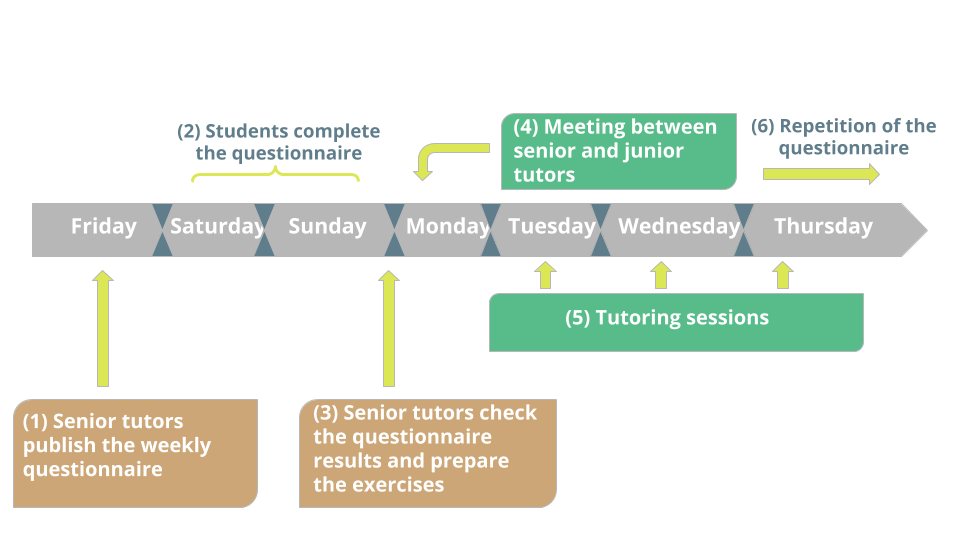}
\caption{Tutoring weekly plan.} \label{fig:week}
\end{center}
\end{figure}

The weekly activities began one month after the beginning of the courses and were organised as follows (see also Figure \ref{fig:week}). (1) Senior tutors prepared a set of multiple choice questions on selected issues from the lectures; these questions were about important concepts or aimed to eliminate typical misconceptions.
(2) Students were invited to answer the weekly questionnaire during the weekend. (3) On the basis of the results, senior tutors prepared specific exercises in order to address common errors.  (4)The exercises were discussed on the following Monday during an online meeting between senior and junior tutors: the aim of the meeting was to inform junior tutors about the students' results in the previous questionnaire in order to outline the expected errors for the proposed exercises. (5)   Each tutoring group had two weekly sessions (scheduled on different days at different times in order to give all the students the possibility to attend the class); in each session the same exercises were used. The interactions with the students during the tutoring sessions took place in the chatroom of the videoconferencing system  and, also, using dedicated Telegram channels. Each session was divided into two parts. In the first part, students had time to go through the proposed exercises on their own, in the second part the exercises were solved by the junior tutors. All tutoring sessions were recorded and, then, made available on the Moodle platform, together with the exercises and their solutions. (6) Students were requested to complete the same questionnaire (with the same questions) after each tutoring session in order to measure the impact of the tutoring activities. 

\end{widetext}

The  tutoring was organised over  8 weeks (plus 1 for revision of mechanics) and the  topics covered the whole content of the Physics I course, in the following sequence:
\begin{itemize}
\item Week 1: Mathematical prerequisites
\item Week 2: Kinematics
\item Week 3: Dynamics
\item Week 4: Conservation laws and collisions
\item Week 5: Rigid body
\item Week 6: Gravitation and electrostatics
\item Revision of Mechanics
\item Week 7: Calorimetry and first law of thermodynamics
\item Week 8: Second law of thermodynamics
\end{itemize} 

The entire tutoring lasted for 9 weeks, with 18 tutoring sessions per group. The whole set of  activities   was recorded both on the Moodle platform and on the Telegram channels: they are currently being  analysed and will be discussed in a forthcoming paper \cite{preparazione}. In this work, we focus on two research questions that are important to evaluate the effectiveness of this tutoring model, both during its progress and  in the final  exams, namely:\\

\noindent \textit{Q1: As a result of the weekly activities performed and of the interactions with the junior tutors, is there an improvement in the results of the questionnaires?}\\

\noindent \textit{Q2: As a result of the participation in the tutoring activities during the semester, is there an improvement in the marks among the students who attended the tutoring, with respect to the average results of all students?}\\

These research questions will be  discussed in the following Section.

%------------------------Section-------------------------
\section{Methodologies and Results}\label{sec:res}
%------------------------Section-------------------------

The average number of students who attended  the tutoring sessions every week, as registered by the junior tutors, was 385: this was the sum of the 9 tutoring groups, which means more or less 43 students per week per group. These data  are aggregate, so the number of presences was not equally distributed in each  group and  the same students did not attend all sessions. Nonetheless these data give an idea of the synchronous interactions which occurred  during the tutoring. Interestingly, the number of asynchronous interactions was far greater than the average number students attending (385): in fact by the end of the summer exam session, we had registered about 9000 students on the Moodle platform. Even though the tutoring ended at the beginning of June, students continued to   use this platform after this date for  the video recordings of the tutoring sessions, for the solutions of the problems proposed and  for repeating the weekly questionnaires.

The following analysis refers to all data collected until the end of September, after the end of the summer and autumn exam sessions. Firstly, we analyse the weekly questionnaires, in order to track improvements in the students' performance.  Subsequently we  compare the exam results of the students who attended the tutoring to those achieved by all students.

%------------------------Section-------------------------
\subsection{Weekly Questionnaires}\label{ssec:weekly}
%------------------------Section-------------------------

\begin{figure}[ht]
\begin{center}
\includegraphics[scale=.40]{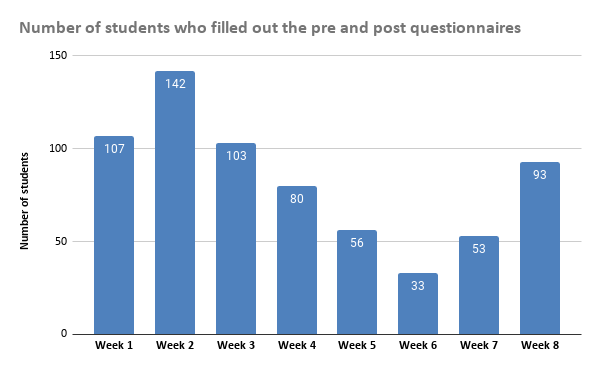}
\caption{Number of students, per week, who carried out both the pre and the post tutoring questionnaires.} \label{fig:involved}
\end{center}
\end{figure}

The weekly questionnaires, one  \textit{before} the tutoring sessions, and one \textit{after} the tutoring sessions, were completed on a voluntary basis. Only one attempt was allowed for both questionnaires. In Figure \ref{fig:involved} we report the number of students who filled out both questionnaires in each week. This number is smaller than the number of students who answered each single questionnaire. For instance, during the first week 20\% of students repeated the test after the tutoring session; in the subsequent weeks the percentages are 32\%, 38\%, 36\%, 34\%, 22\%, 29\% and 23\%.

\begin{figure}[ht]
\begin{center}
\includegraphics[scale=.40]{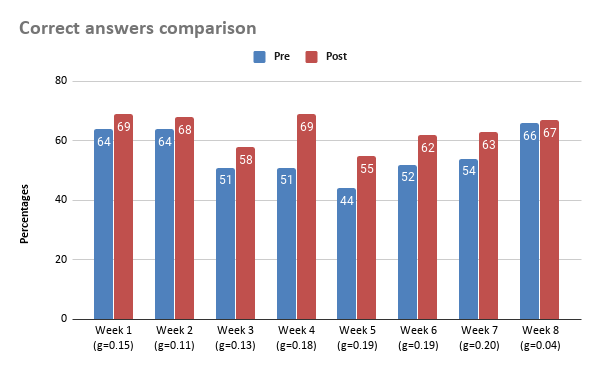}
\caption{Comparison between the percentage of correct answers in the pre and post tutoring questionnaires, together with the corresponding gain ($\mathrm{g}$).} \label{fig:comparison}
\end{center}
\end{figure}

In Figure \ref{fig:comparison} we compare the percentages of correct answers of the pre and post tutoring questionnaires.  To measure the students' performances after the tutoring session, we calculated the Hake normalised gains (reported below each column bar in Figure \ref{fig:comparison}) defined as \cite{hake1998interactive} 
\[
 \mathrm{g}=\frac{P_{f}-P_{i}}{100-P_{i}},
\]
where $P_{i}, P_{f}$ are, respectively, the percentages of correct answers before and after the tutoring sessions. In particular, the gain is 
\begin{itemize}
\item high, if $\mathrm g \geq 0.7$,
\item medium, if $\mathrm  0.7 \geq g \geq 0.3$,
\item low, if $\mathrm g < 0.3$.
\end{itemize}

The gain is defined as the ratio of the actual average gain  to the maximum possible average gain.

%------------------------Section-------------------------
\subsection{Final Exams}\label{ssec:fexa}
%------------------------Section-------------------------

More specific data  about the tutoring were obtained by an online survey administered to all students on the Moodle platform: we collected 1209 answers. The survey covered various areas, such as  the numbers of sessions attended, the most interesting topics for students, the perceived effectiveness of the methods and informations about the interactions with the junior tutors.

\begin{figure}[ht]
\begin{center}
\includegraphics[scale=.4]{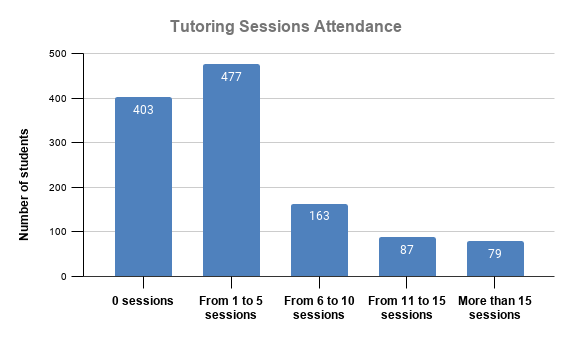}
\caption{Description of the sample analysed: number of sessions attended.} \label{fig:attendance}
\end{center}
\end{figure}

The number of attended sessions is shown in Figure \ref{fig:attendance}. We see that  403 students did not take part in any session: they did not attend the synchronous activities but  they attended the asynchronous ones, such as watching the video recordings, completing the questionnaires, downloading the problems and so on. Hence,  in the following analysis, the sample size is $N=1209-403=806$ students, given by the number of students who attended
 at least one tutoring session. We examined their performances  in the exams of June, July and September. 
 
\begin{widetext}

 \begin{table}[htp]
\begin{center}
\begin{tabular}{|c|c|c|c|c|c|} \hline
Session & Sample  & Quizzes Attempted  & Quizzes Passed  &  Percentage of Quizzes Passed & Average Mark \\  \hline 
\textbf{June}&Tutored Students & 742 &  607 & 81.81 & 10.02  \\
&All Students & 3522 &  2557 & 72.60 & 9.31 \\  \hline
\textbf{July}&Tutored Students & 411 &  354 & 86.13 & 10.04  \\
&All Students & 2636 &  1908 & 72.38 & 8.99 \\  \hline
\textbf{September}& Tutored Students & 195 &  157 & 80.51 & 9.72  \\
&All Students & 1870 &  1325 & 70.86 & 9.02 \\  \hline
\end{tabular}
\end{center}
\caption{Details of June, July and September exams.}\label{tab:jjs}
\end{table}%
 
 In particular, we considered their performance in the first part of the exam (the preliminary test) which consisted of a quiz with 15 multiple choice questions. A minimum of 8 correct answers  was needed to access the second part of the exam, which was an exercise to be solved on-line. A comparison between the results of our sample and those of all students is  carried out below. The (normalized) results of the statistical analysis are reported in Figure \ref{fig:compex}, while the details are in Table \ref{tab:jjs}.

\begin{figure}[ht]
\begin{center}
\includegraphics[scale=.40]{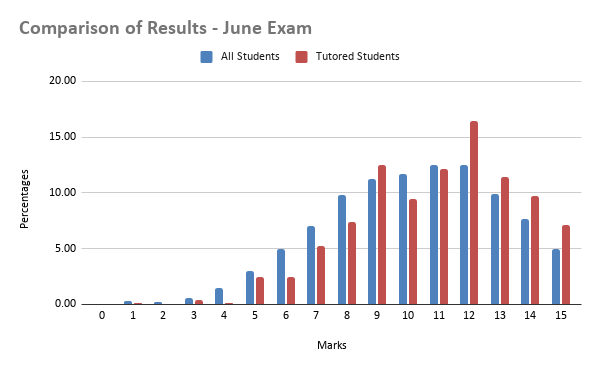}
\includegraphics[scale=.40]{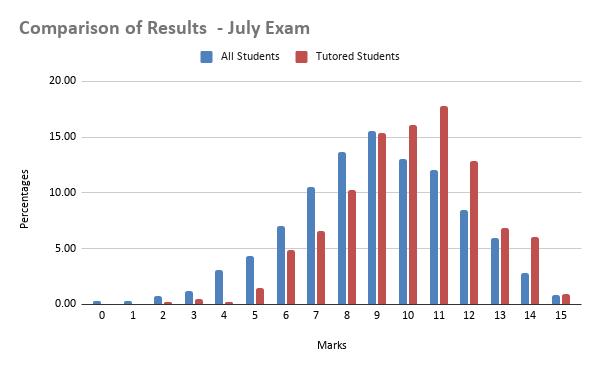}
\includegraphics[scale=.40]{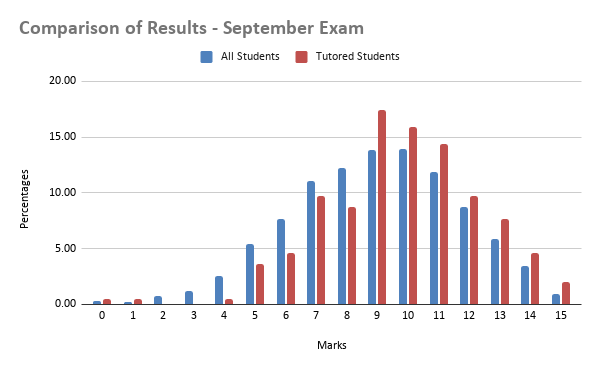}
\caption{Comparison of the students results in June, July and September Physics 1 exams.} \label{fig:compex}
\end{center}
\end{figure}

 As we can see looking at the distributions in Figure \ref{fig:compex} and  Table \ref{tab:jjs}, the average mark in the preliminary multiple choice test is higher for the sample of students who attended the tutoring sessions: this positive trend is evident in all  exam sessions. In order to quantify the statistical significance of the difference between the average marks, we decided to perform a T-test and to evaluate the p-value. In our case the the null hypothesis is: \textit{there is no statistical difference among the average marks of the two samples, their difference is  due to a statistical fluctuation}.

 \begin{table}[htp]
\begin{center}
\begin{tabular}{|c|c|c|c|c|c|} \hline
Session & Sample  & Variance & Standard Deviation &  Average Mark & Sample Size \\  \hline 
\textbf{June}&Tutored Students & 7.43 &  2.73 & 10.02 & 742  \\
&All Students & 8.56 &  2.92 & 9.31 & 3522 \\  \hline
\textbf{July}&Tutored Students & 5.33 &  2.31 & 10.04 & 411  \\
&All Students & 7.27 &  2.69 & 8.99 & 2635 \\  \hline
\textbf{September}& Tutored Students & 6.58 &  2.57 & 9.72 & 195  \\
&All Students & 7.49 &  2.74 & 9.02 & 1870 \\  \hline
\end{tabular}
\end{center}
\caption{Statistical parameters of the two samples.}\label{tab:stat}
\end{table}%
 
\end{widetext}

The T-test was carried out on the basis of the statistical parameters reported in  Table \ref{tab:stat}. In all cases (June, July and September sessions) the resulting p-value is smaller than 0.001, meaning that we have a confidence level bigger that 99.9\%  to reject the null hypothesis. In other words, the different average marks of the two samples are not due to statistical fluctuations.

\begin{figure}[ht]
\begin{center}
\includegraphics[scale=.2]{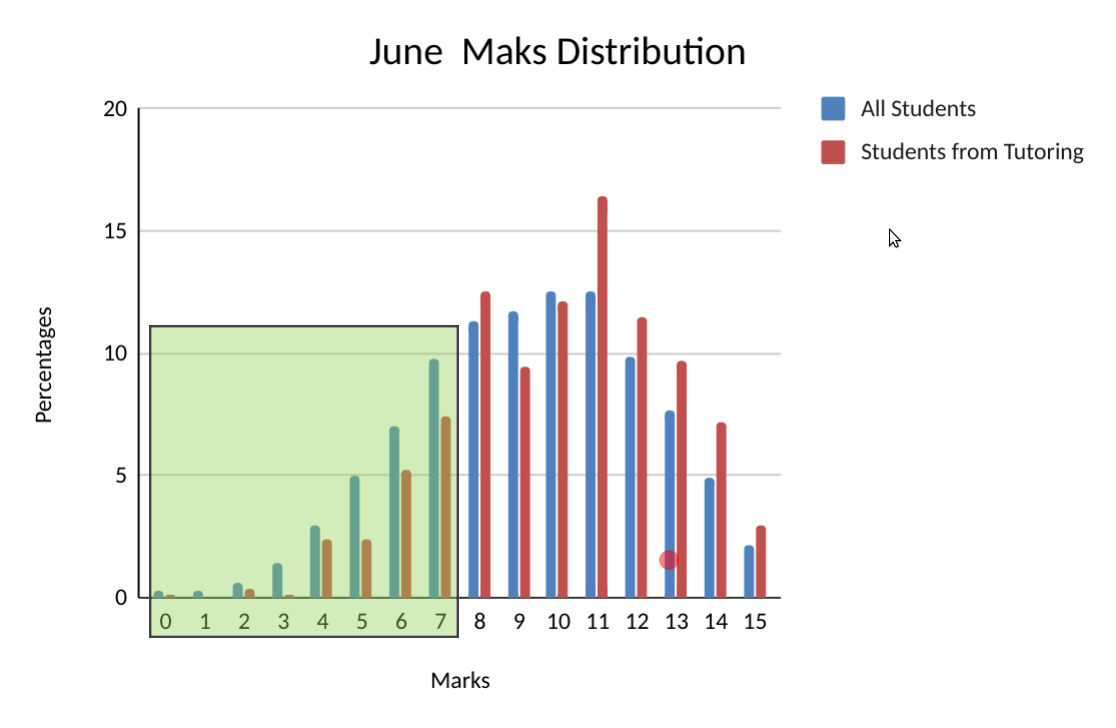}
\includegraphics[scale=.2]{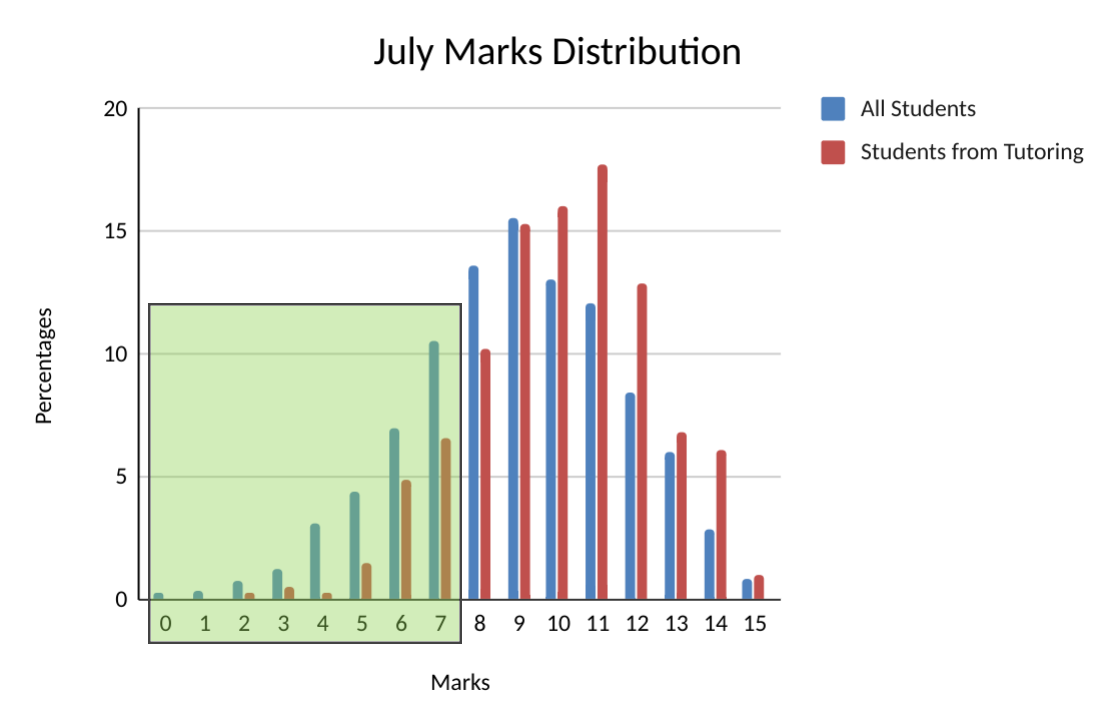}
\includegraphics[scale=.2]{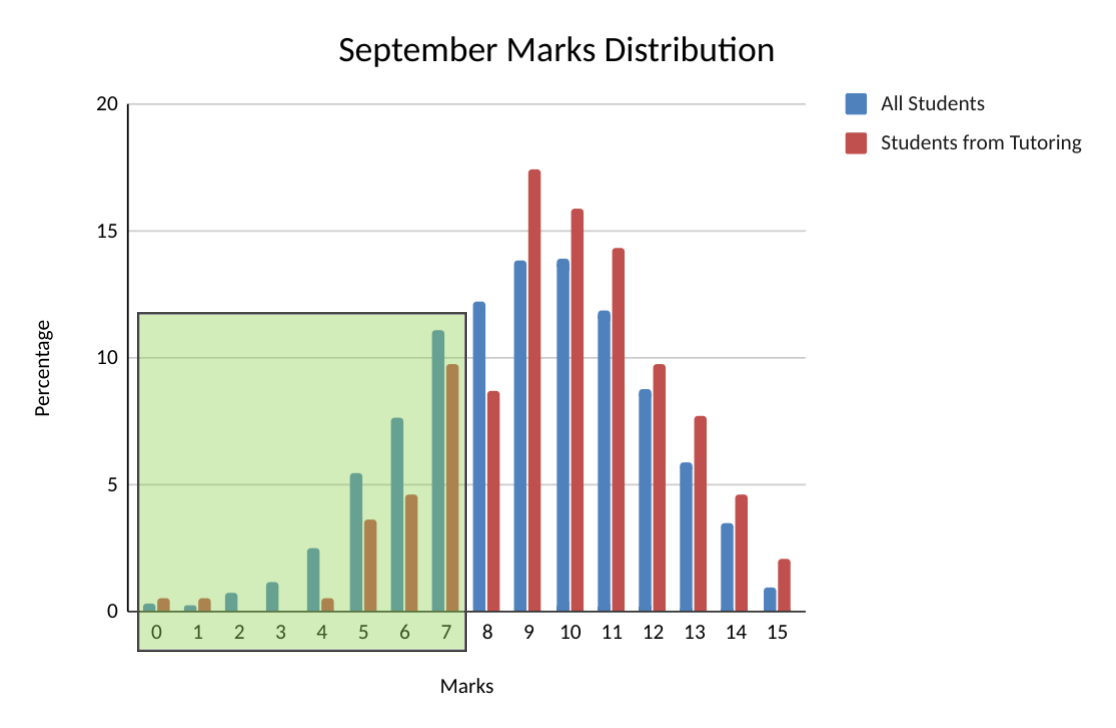}
\caption{Comparison among the total percentage of rejected students in the preliminary test. In all  exams sessions, the  sample of students from Tutoring shows a lowered percentage.} \label{fig:nons}
\end{center}
\end{figure}

From the data in table \ref{tab:stat}, we see that the marks of the students from tutoring are, on average, 0.8 points higher. It is important to emphasise that, even if this result might seem to be negligible, it has significantly decreased the number of student failures (grade $<8$)  in the Physics I course. By comparing the percentages of students with a grade lower than 8, we can evaluate the improvement in the rate of successful students (see Figure \ref{fig:nons}).   In June the percentage of Tutored students not passing the test was lowered by 9\%, in July by 14\%, in September by 10\%. If we apply the average 11\% improvement to the number of students attending the Physics I courses  each year ($\sim 10000$), the tutoring activities would lead to  an increase of about $ \sim 1000$  in the number of successful students in the preliminary test.

%------------------------Section-------------------------
\section{Discussion}\label{sec:disc}
%------------------------Section-------------------------

Despite the low completion rate for the questionnaires, a small improvement in the student marks was evident (see Section \ref{ssec:weekly}). The average weekly attendance was 385 students and in the best case (Week 2),  only 142 students (37 \%) filled out both the pre and post questionnaires. Moreover, if we exclude the Week 1 questionnaire  which focused on mathematical prerequisites, attendance decreased significantly between Week 2 and Week 6, and then increased gradually over the last two weeks. 

This variation in attendance could be due to a possible correlation between the increasing difficulty of the topics and the  time required for study. Therefore, students may have been unable to complete the post questionnaire in due time.
 Furthermore, the  growth in attendance for the last two weeks could have been provoked by the approach of the exam session. 

If we consider the results summarised in Figure \ref{fig:comparison}, according to the Hake scale, the gain is low even if there is an improvement. As reported by \citet{nissen2018comparison} ``Hake adopted this standardizing coefficient because it accounted for the smaller shift in means that could occur in courses with higher pretest means''. On the other hand, this coefficient tends to show a low student improvement when the pretest means are  low. Furthermore, if we look at Figure \ref{fig:comparison}, we see that in our case the mean pretest results are around  50 \%.   However, if pretest means are around 100\%, then improvements are also high.

We obtained better results for the second research question, as reported in Section \ref{ssec:fexa}. In Figure \ref{fig:compex} we see that in all exam sessions the distribution of marks for students who attended the tutoring were higher than average. Moreover,   Table \ref{tab:jjs} shows that the percentage of students who passed the tests is noticeably  greater in the sample of students who attended the tutoring sessions.  The  T-test analysis suggests that the improvements were not due to statistical fluctuations.

Obviously, it is not possible to attribute these positive results entirely to the tutoring intervention. However, we suppose  that students who took part in the tutoring activities were more diligent than the others. Indeed, tutoring gave them  continuous practice and revision of the physics concepts, in addition to what they learned during the lectures. It is widely accepted that  time-on-task deeply influences  learning outcomes \cite{carroll1963model,bloom1974time}. However,  tutoring can also be considered as important, since it gives all students the possibility to revise difficult topics and to reach the required knowledge level  for exam success.

%------------------------Section-------------------------
\section{Conclusions}\label{sec:conc}
%------------------------Section-------------------------

Due to the lockdown provoked by the Covid-19 pandemic, schools and universities  around the world made huge efforts to adopt teaching methods based on distance learning. Social distancing impacted also on laboratory work and tutoring, which have an important role in the peer instruction model. In particular, tutoring models involving advanced students who  meet small groups of students, needed to be redesigned, to cope with the new conditions determined by the pandemic. We have discussed the on line tutoring model developed for the introductory physics courses at Turin Polytechinc University, and analysed its impact on the student learning process, both during the tutoring activities and, at the end, during the exams. In this paper, we have focused on  two research questions to evaluate the effectiveness of the tutor activities, during the semester, by analysing the results of weekly questionnaires (Q1), and at the end of the semester, during the exams in the summer and autumn sessions, by comparing the results of the students who attended the tutoring with the results of all students (Q2).

The results  for Q1 showed that the improvement in the student performances was low, as measured by the Hake scale. However, it is important to emphasise that only a  small number of students involved in the tutoring activities completed both the weekly pre and  post tutoring questionnaires. We suggest that, due to the increasing difficulty in the content of the  lectures, students needed more time to study,  so  they were unable to completethe questionnaires in time. 

The data collected during the June, July and September exam sessions showed that, in the preliminary tests, the distribution of marks for the students who attended the tutoring were better than the overall results; we performed a T-test analysis and showed that the improvement was not due to statistical fluctuations. We also suggest that the time-on-task, due to the activities performed during the tutoring, contributed  to this improvement. Moreover we estimated a 10\% increase in the number of successful students in the preliminary tests due to the tutoring activities.

In conclusion, even though we know that further improvements are needed, for instance to stimulate a greater participation,  we believe that it was important to give the students the possibility of experiencing the peer education model through the tutoring activities, during the lockdown.
Moreover, the improvement in the students' performances were  encouraging and suggests the effectiveness of this tutoring model.

\bibliography{draft_t.bib}

\begin{thebibliography}{21}
\expandafter\ifx\csname natexlab\endcsname\relax\def\natexlab#1{#1}\fi
\expandafter\ifx\csname bibnamefont\endcsname\relax
  \def\bibnamefont#1{#1}\fi
\expandafter\ifx\csname bibfnamefont\endcsname\relax
  \def\bibfnamefont#1{#1}\fi
\expandafter\ifx\csname citenamefont\endcsname\relax
  \def\citenamefont#1{#1}\fi
\expandafter\ifx\csname url\endcsname\relax
  \def\url#1{\texttt{#1}}\fi
\expandafter\ifx\csname urlprefix\endcsname\relax\def\urlprefix{URL }\fi
\providecommand{\bibinfo}[2]{#2}
\providecommand{\eprint}[2][]{\url{#2}}

\bibitem[{\citenamefont{O'Brien}(2020)}]{o2020feynman}
\bibinfo{author}{\bibfnamefont{D.~J.} \bibnamefont{O'Brien}},
  \bibinfo{journal}{arXiv preprint arXiv:2008.07441}  (\bibinfo{year}{2020}).

\bibitem[{\citenamefont{Tan and Chen}(2020)}]{tan2020bringing}
\bibinfo{author}{\bibfnamefont{D.~Y.} \bibnamefont{Tan}} \bibnamefont{and}
  \bibinfo{author}{\bibfnamefont{J.-M.} \bibnamefont{Chen}},
  \bibinfo{journal}{arXiv preprint arXiv:2009.02705}  (\bibinfo{year}{2020}).

\bibitem[{\citenamefont{Dew et~al.}(2020)\citenamefont{Dew, Perry, Ford,
  Nodurft, and Erukhimova}}]{dew2020student}
\bibinfo{author}{\bibfnamefont{M.}~\bibnamefont{Dew}},
  \bibinfo{author}{\bibfnamefont{J.}~\bibnamefont{Perry}},
  \bibinfo{author}{\bibfnamefont{L.}~\bibnamefont{Ford}},
  \bibinfo{author}{\bibfnamefont{D.}~\bibnamefont{Nodurft}}, \bibnamefont{and}
  \bibinfo{author}{\bibfnamefont{T.}~\bibnamefont{Erukhimova}},
  \bibinfo{journal}{arXiv preprint arXiv:2009.11393}  (\bibinfo{year}{2020}).

\bibitem[{\citenamefont{Klein et~al.}(2020)\citenamefont{Klein, Ivanjek,
  Dahlkemper, Jeli{\v{c}}i{\'c}, Geyer, K{\"u}chemann, and
  Susac}}]{klein2020studying}
\bibinfo{author}{\bibfnamefont{P.}~\bibnamefont{Klein}},
  \bibinfo{author}{\bibfnamefont{L.}~\bibnamefont{Ivanjek}},
  \bibinfo{author}{\bibfnamefont{M.~N.} \bibnamefont{Dahlkemper}},
  \bibinfo{author}{\bibfnamefont{K.}~\bibnamefont{Jeli{\v{c}}i{\'c}}},
  \bibinfo{author}{\bibfnamefont{M.-A.} \bibnamefont{Geyer}},
  \bibinfo{author}{\bibfnamefont{S.}~\bibnamefont{K{\"u}chemann}},
  \bibnamefont{and} \bibinfo{author}{\bibfnamefont{A.}~\bibnamefont{Susac}},
  \bibinfo{journal}{arXiv preprint arXiv:2010.05622}  (\bibinfo{year}{2020}).

\bibitem[{\citenamefont{Moore}(1973)}]{moore1973toward}
\bibinfo{author}{\bibfnamefont{M.~G.} \bibnamefont{Moore}},
  \bibinfo{journal}{The Journal of Higher Education}
  \textbf{\bibinfo{volume}{44}}, \bibinfo{pages}{661} (\bibinfo{year}{1973}).

\bibitem[{\citenamefont{Dietrich et~al.}(2020)\citenamefont{Dietrich,
  Kentheswaran, Ahmadi, Teychen{\'e}, Bessi{\`e}re, Alfenore, Laborie, Bastoul,
  Loubi{\`e}re, Guigui et~al.}}]{dietrich2020attempts}
\bibinfo{author}{\bibfnamefont{N.}~\bibnamefont{Dietrich}},
  \bibinfo{author}{\bibfnamefont{K.}~\bibnamefont{Kentheswaran}},
  \bibinfo{author}{\bibfnamefont{A.}~\bibnamefont{Ahmadi}},
  \bibinfo{author}{\bibfnamefont{J.}~\bibnamefont{Teychen{\'e}}},
  \bibinfo{author}{\bibfnamefont{Y.}~\bibnamefont{Bessi{\`e}re}},
  \bibinfo{author}{\bibfnamefont{S.}~\bibnamefont{Alfenore}},
  \bibinfo{author}{\bibfnamefont{S.}~\bibnamefont{Laborie}},
  \bibinfo{author}{\bibfnamefont{D.}~\bibnamefont{Bastoul}},
  \bibinfo{author}{\bibfnamefont{K.}~\bibnamefont{Loubi{\`e}re}},
  \bibinfo{author}{\bibfnamefont{C.}~\bibnamefont{Guigui}},
  \bibnamefont{et~al.}, \bibinfo{journal}{Journal of Chemical Education}
  (\bibinfo{year}{2020}).

\bibitem[{\citenamefont{Damon}(1984)}]{damon1984peer}
\bibinfo{author}{\bibfnamefont{W.}~\bibnamefont{Damon}},
  \bibinfo{journal}{Journal of applied developmental psychology}
  \textbf{\bibinfo{volume}{5}}, \bibinfo{pages}{331} (\bibinfo{year}{1984}).

\bibitem[{\citenamefont{Lasry et~al.}(2008)\citenamefont{Lasry, Mazur, and
  Watkins}}]{lasry2008peer}
\bibinfo{author}{\bibfnamefont{N.}~\bibnamefont{Lasry}},
  \bibinfo{author}{\bibfnamefont{E.}~\bibnamefont{Mazur}}, \bibnamefont{and}
  \bibinfo{author}{\bibfnamefont{J.}~\bibnamefont{Watkins}},
  \bibinfo{journal}{American journal of Physics} \textbf{\bibinfo{volume}{76}},
  \bibinfo{pages}{1066} (\bibinfo{year}{2008}).

\bibitem[{\citenamefont{Mazur}(2013)}]{mazur2013peer}
\bibinfo{author}{\bibfnamefont{E.}~\bibnamefont{Mazur}},
  \emph{\bibinfo{title}{Peer Instruction: Pearson New International Edition: A
  User's Manual}}, Always learning (\bibinfo{publisher}{Pearson Higher
  Education \& Professional Group}, \bibinfo{year}{2013}), ISBN
  \bibinfo{isbn}{9781292039701},
  \urlprefix\url{https://books.google.it/books?id=OReioAEACAAJ}.

\bibitem[{\citenamefont{Crouch and Mazur}(2001)}]{crouch2001peer}
\bibinfo{author}{\bibfnamefont{C.~H.} \bibnamefont{Crouch}} \bibnamefont{and}
  \bibinfo{author}{\bibfnamefont{E.}~\bibnamefont{Mazur}},
  \bibinfo{journal}{American journal of physics} \textbf{\bibinfo{volume}{69}},
  \bibinfo{pages}{970} (\bibinfo{year}{2001}).

\bibitem[{\citenamefont{Boud et~al.}(2014)\citenamefont{Boud, Cohen, and
  Sampson}}]{boud2014peer}
\bibinfo{author}{\bibfnamefont{D.}~\bibnamefont{Boud}},
  \bibinfo{author}{\bibfnamefont{R.}~\bibnamefont{Cohen}}, \bibnamefont{and}
  \bibinfo{author}{\bibfnamefont{J.}~\bibnamefont{Sampson}},
  \emph{\bibinfo{title}{Peer learning in higher education: Learning from and
  with each other}} (\bibinfo{publisher}{Routledge}, \bibinfo{year}{2014}).

\bibitem[{\citenamefont{Collings et~al.}(2014)\citenamefont{Collings, Swanson,
  and Watkins}}]{collings2014impact}
\bibinfo{author}{\bibfnamefont{R.}~\bibnamefont{Collings}},
  \bibinfo{author}{\bibfnamefont{V.}~\bibnamefont{Swanson}}, \bibnamefont{and}
  \bibinfo{author}{\bibfnamefont{R.}~\bibnamefont{Watkins}},
  \bibinfo{journal}{Higher Education} \textbf{\bibinfo{volume}{68}},
  \bibinfo{pages}{927} (\bibinfo{year}{2014}).

\bibitem[{\citenamefont{Sneddon}(1759)}]{sneddon1759attitudes}
\bibinfo{author}{\bibfnamefont{P.~H.} \bibnamefont{Sneddon}},
  \bibinfo{journal}{Journal of Learning Development in Higher Education ISSN}
  p. \bibinfo{pages}{667X} (\bibinfo{year}{1759}).

\bibitem[{\citenamefont{Pols}(2020)}]{pols2020physics}
\bibinfo{author}{\bibfnamefont{F.}~\bibnamefont{Pols}}, \bibinfo{journal}{The
  Electronic Journal for Research in Science \& Mathematics Education}
  \textbf{\bibinfo{volume}{24}}, \bibinfo{pages}{172} (\bibinfo{year}{2020}).

\bibitem[{\citenamefont{Bradbury and Pols}(2020)}]{bradbury2020pandemic}
\bibinfo{author}{\bibfnamefont{F.}~\bibnamefont{Bradbury}} \bibnamefont{and}
  \bibinfo{author}{\bibfnamefont{C.}~\bibnamefont{Pols}},
  \bibinfo{journal}{arXiv preprint arXiv:2006.06881}  (\bibinfo{year}{2020}).

\bibitem[{\citenamefont{Otero et~al.}(2010)\citenamefont{Otero, Pollock, and
  Finkelstein}}]{otero2010physics}
\bibinfo{author}{\bibfnamefont{V.}~\bibnamefont{Otero}},
  \bibinfo{author}{\bibfnamefont{S.}~\bibnamefont{Pollock}}, \bibnamefont{and}
  \bibinfo{author}{\bibfnamefont{N.}~\bibnamefont{Finkelstein}},
  \bibinfo{journal}{American Journal of Physics} \textbf{\bibinfo{volume}{78}},
  \bibinfo{pages}{1218} (\bibinfo{year}{2010}).

\bibitem[{\citenamefont{Galante and Ruggiero}(2020)}]{preparazione}
\bibinfo{author}{\bibfnamefont{L.}~\bibnamefont{Galante}} \bibnamefont{and}
  \bibinfo{author}{\bibfnamefont{M.~L.} \bibnamefont{Ruggiero}},
  \bibinfo{journal}{in preparation}  (\bibinfo{year}{2020}).

\bibitem[{\citenamefont{Hake}(1998)}]{hake1998interactive}
\bibinfo{author}{\bibfnamefont{R.~R.} \bibnamefont{Hake}},
  \bibinfo{journal}{American journal of Physics} \textbf{\bibinfo{volume}{66}},
  \bibinfo{pages}{64} (\bibinfo{year}{1998}).

\bibitem[{\citenamefont{Nissen et~al.}(2018)\citenamefont{Nissen, Talbot,
  Thompson, and Van~Dusen}}]{nissen2018comparison}
\bibinfo{author}{\bibfnamefont{J.~M.} \bibnamefont{Nissen}},
  \bibinfo{author}{\bibfnamefont{R.~M.} \bibnamefont{Talbot}},
  \bibinfo{author}{\bibfnamefont{A.~N.} \bibnamefont{Thompson}},
  \bibnamefont{and}
  \bibinfo{author}{\bibfnamefont{B.}~\bibnamefont{Van~Dusen}},
  \bibinfo{journal}{Physical Review Physics Education Research}
  \textbf{\bibinfo{volume}{14}}, \bibinfo{pages}{010115}
  (\bibinfo{year}{2018}).

\bibitem[{\citenamefont{Carroll}(1963)}]{carroll1963model}
\bibinfo{author}{\bibfnamefont{J.~B.} \bibnamefont{Carroll}},
  \bibinfo{journal}{Teachers college record}  (\bibinfo{year}{1963}).

\bibitem[{\citenamefont{Bloom}(1974)}]{bloom1974time}
\bibinfo{author}{\bibfnamefont{B.~S.} \bibnamefont{Bloom}},
  \bibinfo{journal}{American psychologist} \textbf{\bibinfo{volume}{29}},
  \bibinfo{pages}{682} (\bibinfo{year}{1974}).

\end{thebibliography}


\begin{thebibliography}{200}


\end{thebibliography}

\end{document}